\documentclass[useAMS,usenatbib]{mn2e}
\usepackage{graphicx,natbib}
\newcommand{\Msun}{\mbox{\rm M$_{\odot}$}}
\def\change#1{{\it #1}}

\title[Origin of volatiles in the Main Belt]{Origin of volatiles in the Main Belt}

  \author[Mousis et al.]{O. Mousis$^1$\thanks{E-mail:olivier.mousis@obs-besancon.fr}, Y. Alibert$^{2,1}$, D. Hestroffer$^3$, U. Marboeuf$^1$, C. Dumas$^4$, B. Carry$^4$,
\newauthor  
J. Horner$^5$ and F. Selsis$^6$\\
$^1$ Institut UTINAM, CNRS-UMR 6213, Observatoire de Besan\c{c}on, BP 1615, 25010 Besan\c{c}on Cedex, France\\
$^2$ Physikalisches Institut, University of Bern, Sidlerstrasse 5, CH-3012 Bern, Switzerland\\
$^3$ IMCCE, CNRS-UMR 8028, Observatoire de Paris, 77 Av. Denfert Rochereau 75014 Paris, France \\
$^4$ ESO, Alonso de Cordova 3107, Vitacura, Santiago de Chile, Chile \\
$^5$ Astronomy Group, Open University, Walton Hall, Milton Keynes MK76AA, United Kingdom \\
$^6$  C.R.A.L, Ecole normale sup\'{e}rieure, 46 all\'{e}e d'Italie, 69007 Lyon, France}

\date{Accepted ??. Received ??; in original form ??}

\begin{document}

\pagerange{\pageref{firstpage}--\pageref{lastpage}} \pubyear{2007}

\label{firstpage}

\maketitle
\begin{abstract}
We propose a scenario for the formation of the Main Belt in which asteroids incorporated icy particles formed in the outer Solar Nebula. We calculate the composition of icy planetesimals formed beyond a heliocentric distance of 5 AU in the nebula by assuming that the abundances of all elements, in particular that of oxygen, are solar. As a result, we show that ices formed in the outer Solar Nebula are composed of a mix of clathrate hydrates, hydrates formed above 50 K and pure condensates produced at lower temperatures. We then consider the inward migration of solids initially produced in the outer Solar Nebula and show that a significant fraction may have drifted to the current position of the Main Belt without encountering temperature and pressure conditions high enough to vaporize the ices they contain. We propose that, through the detection and identification of initially buried ices revealed by recent impacts on the surfaces of asteroids, it could
be possible to infer the thermodynamic conditions that were present within the Solar Nebula during the accretion of these bodies, and during the inward migration of icy planetesimals. We also investigate the potential influence that the incorporation of ices in asteroids may have on their porosities and densities. In particular, we show how the presence of ices reduces the value of the bulk density of a given body, and consequently modifies its macro-porosity from that which would be expected from a given
taxonomic type.
\end{abstract}


\section{Introduction}

In recent years, some objects within the Main Belt of asteroids have been found to display cometary characteristics (Hsieh \& Jewitt 2006). Objects such as 133P/Elst-Pizarro, P/2005 U1 and 118401 (1999 RE$_{70}$) occupy orbits that are entirely decoupled from Jupiter within the Main Belt, and are probably bodies that have undergone a recent collision, revealing previously buried volatile material, and leading to the observed dusty outgassing.  In addition, present-day surface water ice and possible water sublimation have been reported on Ceres (Lebofsky et al. 1981; A'Hearn \& Feldman 1992; Vernazza et al. 2005). This is consistent with recent Hubble Space Telescope (HST) observations which suggest that Ceres' shape is the result of the dwarf planet consisting of a rocky core surrounded by an ice-rich mantle (Thomas et al. 2005) - an idea in agreement with several thermal evolution scenarios (McCord \& Sotin 2005) that suggest that the ice content of the asteroid is between 17\% and 27\%, by mass. These observations are supported by the evidence of hydrated minerals in meteorites which provide samples of rock from asteroids in the Main Belt. Most of these minerals formed as a result of water ice accreting with the chondritic meteorite parent bodies, melting, and driving aqueous alteration reactions (Clayton \& Mayeda 1996; Jewitt et al. 2007).

It seems likely, then, that some objects in the asteroid belt have incorporated significant amounts of water ice (and possibly other volatiles) during their formation in the early stages of the solar
System. These bodies would have incorporated icy particles\footnote{By icy particles is meant planetesimals composed of a mix of ices and rocks.} coming from the outer nebula that survived their inward drift due to gas-drag through the disk (Mousis \& Alibert 2005 -- hereafter MA05). The volatile fraction incorporated in this manner could vary depending on the inward flux of icy planetesimals from the external region and the heliocentric location of the asteroid, together with the density of the proto-solar nebula. Given that the current asteroid belt lies closer to the Sun than the ``snow-line'', postulated to lie at around 5 AU in the Solar Nebula, these results are a little unexpected. In this context, understanding how volatiles were incorporated into the asteroids is therefore important, not only for the study of the asteroids themselves, but also for our understanding of the processes by which the solar system came into being.

To this end, MA05 studied the possibility of determining the nature and composition of the ices which were incorporated into Ceres. They used a time dependant model of the Solar Nebula and showed that icy particles of sizes between 0.1 and 10 metres could drift from heliocentric distances greater than 5 AU to the present location of Ceres without encountering temperatures or pressures high enough to vaporise the ices within. The authors then suggested that ices produced in the outer Solar Nebula were transported inwards to become incorporated in the solids which accreted to form Ceres. 

The present work aims to improve upon the calculation detailed in MA05, along with expanding the results to involve the entire asteroid belt, rather than just its largest member. In particular, MA05 postulated that all volatiles, except CO$_2$\footnote{CO$_2$ is the only major volatile species which does not form a clathrate hydrate in the Solar Nebula because it condenses as a pure ice prior to being trapped by water.}, were trapped by water in the form of hydrates or clathrate hydrates in the outer solar
nebula. This assumption was supported by the work of Hersant et al. (2001) who estimated that Jupiter was formed at temperatures higher than $\sim$40 -- 50 K. The accretion of ices in the form of hydrates and clathrate hydrates was thus required during the formation of the planet in order to explain the volatile enrichments observed in its atmosphere\footnote{The abundances of volatile species in Jupiter's atmosphere have been measured using the mass spectrometer on board the {\it Galileo} probe. These measurements reveal that the giant planet's atmosphere is enriched by a factor of $\sim$3 in Ar, Kr, Xe, C, N, and S compared to the solar abundances (Owen et al. 1999).} (Gautier et al. 2001a,b). Indeed, since these ices are usually formed at temperatures higher than that reached by the nebula at the time of Jupiter's completion, as defined by Hersant et al. (2001), they can be incorporated in the planetesimals accreted by the giant planet during its growth. However, the amount of water that would be required in the nebula to trap all these volatiles as hydrates and clathrate hydrates exceeds that derived from the solar oxygen abundance. Therefore, MA05 made the {\it ad hoc} hypothesis that oxygen was ``oversolar'' in the gas-phase in order to provide enough available water in the Solar Nebula\footnote{ The oxygen abundance required for to allow the trapping of all volatile species in the form of hydrates or clathrate hydrates is $\sim$1.9 times the solar abundance, with CO$_2$:CO:CH$_4$ = 1:1:1 and N$_2$:NH$_3$ = 1:1 (the nominal nebula gas phase ratios used in this work).}. Additionally, Hersant et al. (2001) only used an evolutionary Solar Nebula model to derive the disk's temperature at the time when the mass of Jupiter's feeding zone was equal to that of the gas in its current envelope. They thus neglected many important effects such as the influence of protoplanet formation on the structure of the disk (e.g. Fig. 2 of Alibert et al. 2004). However, recent giant planet core-accretion formation models that include migration, disk evolution, such as that proposed by Alibert et al. (2004), have shown that the disk's temperature can be as low as $\sim$10 -- 20 K at the end of Jovian formation. This implies that Jupiter itself can accrete ices during its formation that were produced at temperatures lower than those required for clathration. As a result, no extra water is required in the nebula to allow all the volatile species to be trapped in clathrate hydrates, and the oversolar oxygen abundance condition in the nebula can be relaxed.

In Section 2, we calculate the composition of ices produced in the outer Solar Nebula under the assumption that the abundances of all elements, in particular that of oxygen, are solar. In Section 3, we consider the inward migration of particles produced at various locations in the nebula, and at different
times. This allows us to examine whether some planetesimals formed in the outer Solar Nebula may have drifted to the current position of the Main Belt without encountering temperature and pressure conditions high enough to vaporize the ices they contain. In Section 4, we examine the uncertainties in the determination of the physical properties of asteroids. We also investigate the potential influence that the incorporation of ices in these objects may have on their porosities and densities. Section 5 is
devoted to summary and discussion.

\section{Composition of icy planetesimals formed in the outer Solar Nebula}

\subsection{Initial gas-phase conditions in the nebula}

In this Section, we aim to estimate the composition of icy particles produced in the region of giant planet formation, prior to the dissipation of the nebula, which then migrated into the forming Main Belt. The calculations have been carried out in a manner consistent with the formation of Jupiter with a realistic primordial volatile composition. This implies that the icy planetesimals that drifted inwards to the formation zone of the asteroids shared the same composition as those accreted by proto-Jupiter. 

We then assume that the gas-phase abundances of elements are solar (e.g. Table \ref{lodders}) and that O, C, and N exist only in the form of H$_2$O, CO$_2$, CO, CH$_4$, N$_2$, and NH$_3$. S is only present as H$_2$S and other refractory sulfur components (Pasek et al. 2005). The gas-phase molecular ratios in the Solar Nebula are presumed to derive directly from that in interstellar ices (Mousis et al. 2002) and, in some cases, from the consideration of catalytic effects that might affect these ratios in the nebula. Thus, although the CO:CH$_4$ ratio is typically around 10:1 in the ISM (Allamandola et al. 1999), this can be revised down in the Solar Nebula gas-phase since some additional CH$_4$ may be
introduced through conversions of both CO and CO$_2$, as a result of the presence of catalytically active regions in the disk (Kress \& Tielens 2001; Sekine et al. 2005). We therefore adopt CO:CH$_4$ = 1:1 in the vapor phase of the nebula. Moreover, CO$_2$ should initially be present in the gas-phase of the nebula, with CO$_2$:CO = 1:1 -- 4:1, a range of values that covers the ISM measurements (Gibb et al. 2004). The value of the N$_2$:NH$_3$ ratio is quite uncertain, although current chemical models of the ISM predict that molecular nitrogen should be much more abundant than ammonia (Irvine \& Knacke 1989). On the other hand, the N$_2$:NH$_3$ ratio may have been much lower in the Solar Nebula since the conversion of N$_2$ into NH$_3$ can be accelerated by the catalytic effect of local Fe grains (Lewis \& Prinn 1980; Fegley 2000). In all our following calculations, we consider N$_2$:NH$_3$ = 1:1 as the nominal ratio for these two molecules in the nebula gas-phase.

\begin{table}
\caption[]{Gas phase abundances (molar mixing ratio with respect to H$_2$) of major species in the Solar Nebula from Lodders (2003) with CO$_2$:CO:CH$_4$ = 1:1:1 and N$_2$:NH$_3$ = 1:1.}
\begin{center}
\begin{tabular}{lclc}
\hline
\hline
\noalign{\smallskip}
Species X &  X/H$_2$  & species X  & X/H$_2$ \\	
\noalign{\smallskip}
\hline
\hline
\noalign{\smallskip}
O       	&  1.16.10$^{-3}$ 	&   N$_2$    	&  5.33.10$^{-5}$    \\
C	      	&   5.82.10$^{-4}$    	&   NH$_3$   	& 5.33.10$^{-5}$   \\
N       	&   1.60.10$^{-4}$    	&   S        		&   3.66.10$^{-5}$   \\
H$_2$O   	&   5.78.10$^{-4}$   	&   Ar       		& 8.43.10$^{-6}$   \\
CO$_2$   	&  1.94.10$^{-4}$    	&   Kr       		&  4.54.10$^{-9}$   \\	
CO       	&   1.94.10$^{-4}$   	&   Xe       		&   4.44.10$^{-10}$   \\		   
CH$_4$   	&  1.94.10$^{-4}$   	&             		&                      \\ 	  
\hline
\end{tabular}
\end{center}
\label{lodders}
\end{table}

\subsection{Formation sequence of the different ices in the outer nebula}

As a result of the adoption of a solar gas-phase oxygen abundance, we show here that ices formed at distances greater than 5 AU in the cooling Solar Nebula are a mix of clathrate hydrates and hydrates
formed above 50 K, and pure condensates primarily produced at lower temperatures (but still greater than $\sim$20 K). We note that this hypothesis is supported by the recent work of Mousis \& Marboeuf (2006), who showed that the abundances of volatiles observed in the envelopes of Jupiter and Saturn could be reproduced by using a solar abundance for all elements in the Solar Nebula, and that the calculated amount of heavy elements remains in agreement with internal structure models. The clathration and hydratation processes result from the presence of available crystalline water ice
at the time of volatile-trapping in the Solar Nebula. This statement seems justified when one considers that current scenarios of the formation of the Solar Nebula suggest that most ices falling from the presolar cloud onto the disk were vaporised upon entering the early nebula. Following Chick and Cassen (1997), H$_2$O ice is initially vaporized at locations in the inner 30 AU of the Solar Nebula. With time, the decrease of temperature and pressure led to conditions which allowed the water to re-condense, forming microscopic crystalline ices (Kouchi et al. 1994, Mousis et al. 2000). Once formed, the different ices agglomerated and were incorporated into the growing and drifting particles.

\begin{figure}
\resizebox{\hsize}{!}{\includegraphics[angle=-90]{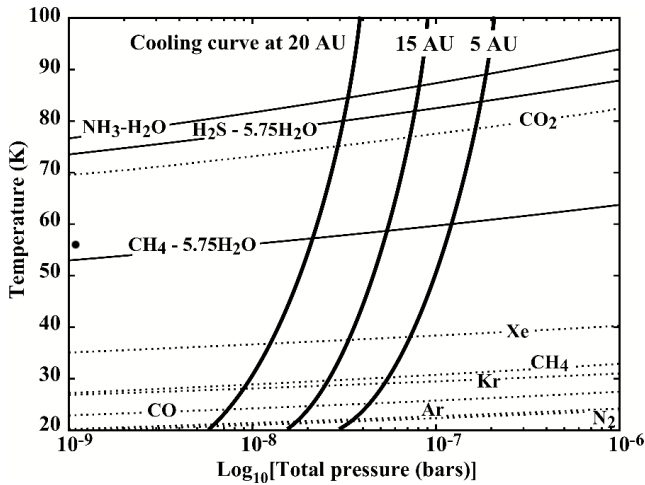}}
\caption{Stability curves for hydrates, clathrate hydrates (solid lines) and pure condensates (dotted lines), together with evolutionary tracks for the nominal protoplanetary disk model at heliocentric distances of 5, 15 and 20 AU. These evolutionary tracks correspond to the thermodynamic pathway
followed by the disk during its cooling at the considered distances to the Sun (the evolution of the disk proceeds from high to low temperatures). The abundances of elements are solar, with CO$_2$:CO:CH$_4$~=~1:1:1 and N$_2$:NH$_3$~=~1:1 in the vapor phase. Species remain in the vapor phase as long as they stay in the domains located above the curves of stability. The label -5.75H$_2$O designates the clathrate hydrate of species X and NH$_3$-H$_2$O corresponds to ammonia hydrate.}
\label{cool_curve}
\end{figure}

The process by which volatiles are trapped, illustrated in Fig. \ref{cool_curve}, is calculated using the stability curves of hydrates, clathrate hydrates and pure condensates, and the tracks detailing the evolution of temperature and pressure at heliocentric distances of 5, 15 and 20 AU within the solar
nebula. These evolutionary tracks are derived from the $\alpha$-turbulent accretion disk model described in next section. The stability curves of hydrates and clathrate hydrates derive from Lunine
\& Stevenson (1985)'s compilation of laboratory data, from which data is available at relatively low temperature and pressure. Their equations, given by Hersant et al. (2004), are of the form $ln~P$ =
$A$/$T$ + $B$, where $P$ and $T$ are the partial pressure (bars) and the temperature (K) of the considered species, respectively. Table \ref{hersant} shows the values of constants $A$ and $B$
determined by Hersant et al. (2004) from their fits to laboratory measurements. On the other hand, the stability curves of pure condensates used in our calculations derive from the compilation of laboratory data given in the CRC Handbook of chemistry and physics (Lide 2002). Their equations are of the form $log~P$ = $A$/$T$ + $B$, where $P$ and $T$ have the same units as mentioned above. Table \ref{pures} also gives the constants $A$ and $B$ as determined by our own fits to laboratory measurements. The cooling curve intercepts the stability curves of the different ices at some given temperature, pressure and surface density conditions (see Table \ref{cond}). For each ice considered, the domain of stability is the region located below its corresponding stability curve. The clathration process stops when no more crystalline water ice is available to trap the volatile species. For example, if one assumes CO$_2$:CO:CH$_4$~=~1:1:1 and N$_2$:NH$_3$~=~1:1 in the gas-phase, then the NH$_3$ and H$_2$S are entirely trapped by the water, as hydrates of NH$_3$ and clathrate
hydrates of H$_2$S, while only approximately half of the CH$_4$ is trapped as a clathrate hydrate in the Solar Nebula. The remaining CH$_4$, as well as Xe, Kr, CO, Ar and N$_2$, whose clathration processes normally occur at higher temperatures, remain in the vapor phase until the Solar Nebula cools enough to allow the formation of pure condensates (roughly between 20 and 30 K in the gas-phase conditions assumed for the Solar Nebula; see Fig. \ref{cool_curve}). Note that, during the cooling of the Solar Nebula, CO$_2$ is the only species that crystallizes as a pure condensate prior to being trapped by water to form a clathrate hydrate. Hence, we assume here that solid CO$_2$ is the only existing
condensed form of CO$_2$ in these environments.

\begin{table}
\caption[]{Parameters of the stability curves of the considered clathrate hydrates (reproduced from Hersant et al. 2004). Their equations are of the form ln~$P$ = $A$/$T$ + $B$, where $P$ and $T$ are the partial pressure (bars) and the temperature (K) of the considered species, respectively. $A$ is in K and $B$ is dimensionless.}
\begin{center}
\begin{tabular}{lcc}
\hline
\hline
\noalign{\smallskip}
Species &  A & B  \\
\noalign{\smallskip}
\hline
\noalign{\smallskip}
CH$_4$     	&    	-2161.81	    	& 	11.1249  \\
CO         		&    	-1685.54      	&   	10.9946  \\
N$_2$      		&    	-1677.62      	& 	11.1919  \\
NH$_3$     	&    	-2878.23      	& 	8.00205   \\
H$_2$S	   	&    	-3111.02       	&  	11.3801 \\
Ar         		&    	-1481.78       	&  	9.95523  \\
Kr         		&    	-1987.5       	&   	9.99046 \\
Xe         		&   	-2899.18       	&   	11.0354  \\	
\hline
\end{tabular}
\end{center}
\label{hersant}
\end{table}

\begin{table}
\caption[]{Parameters of the stability curves of the considered pure condensates. Their equations are of the form log~$P$ = $A$/$T$ + $B$, where $P$ and $T$ are the partial pressure (bars) and the temperature (K) of the considered species, respectively. $A$ is in K and $B$ is dimensionless.}
\begin{center}
\begin{tabular}{lcc}
\hline
\hline
\noalign{\smallskip}
Species &  A & B  \\
\noalign{\smallskip}
\hline
\noalign{\smallskip}
CH$_4$   		&  	-475.61	        	& 4.2831  \\
CO   			&    	-411.24  		&   5.2426  \\
CO$_2$    	&   	-1365.9	    	&  7.0248  \\
N$_2$   		&    	-360.07     	& 4.7459  \\
NH$_3$  		&    	-1565.0  		& 6.7883   \\
H$_2$S	   	&   	-1153.70		&  5.5007 \\
Ar 			&   	-369.90		&  4.1862  \\
Kr 			&    	-603.46      	&   5.1060 \\
Xe  			&   	-819.28    		&   4.9881  \\	
\hline
\end{tabular}
\end{center}
\label{pures}
\end{table}
    
\begin{table}
\caption[]{Temperature $T$(K), total gas pressure $P$(bars) and surface density $\Sigma$(g/cm$^2$) conditions at which different ices form in the Solar Nebula at heliocentric distance of 5 AU. The gas-phase conditions considered here are CO$_2$:CO:CH$_4$ = 1:1:1 and N$_2$:NH$_3$ = 1:1.}
\begin{center}
\begin{tabular}{lccc}
\hline
\hline
\noalign{\smallskip}
Ices &  $T$(K)  & $P$(bar)  & $\Sigma$(g/cm$^2$) \\	
\noalign{\smallskip}
\hline
\noalign{\smallskip}
H$_2$O				&	 157.5	&   3.4 $\times$ 10$^{-7}$		& 677.4		\\
NH$_3$-H$_2$O  		&    89.1	  	&   1.9 $\times$ 10$^{-7}$    	& 498.0  		\\
H$_2$S-5.75H$_2$O	&    84.0  		&    1.8 $\times$ 10$^{-7}$   	& 482.6		\\
CO$_2$   				&    78.0	     	&   1.6 $\times$ 10$^{-7}$       	& 463.8      	\\
CH$_4$-5.75H$_2$O   	&    60.9	   	&   1.3 $\times$ 10$^{-7}$    	& 404.3   		\\
Xe       				&    38.2     	&    7.6 $\times$ 10$^{-8}$    	&  301.1  		\\	
CH$_4$   				&    31.3    		&    6.0 $\times$ 10$^{-8}$     	&  259.1		\\
Kr       				&    29.2	     	&   5.4 $\times$ 10$^{-8}$    	&  244.4   		\\
CO       				&    25.3	    	&   4.5 $\times$ 10$^{-8}$    	&  213.9   		 \\
Ar       				&    22.2	    	&   3.6 $\times$ 10$^{-8}$    	&   185.6     	 \\	
N$_2$    				&    21.8     	&   3.5 $\times$ 10$^{-8}$   	&   181.7  		\\    
\hline
\end{tabular}
\end{center}
\label{cond}
\end{table}

\subsection{Composition of icy planetesimals}
\label{icy}

 Using the trapping/formation conditions of the different ices calculated at a given heliocentric distance in the outer nebula (e.g. Table \ref{cond}), and knowing their gas-phase abundances, one can estimate their mass ratios with respect to H$_2$O in the accreting planetesimals. Indeed, the volatile, $i$, to water mass ratio in these planetesimals is determined by the relation given by Mousis \& Gautier (2004):         

\begin{equation}
{m_i = \frac{X_i}{X_{H_2O}} \frac{\Sigma(R; T_i, P_i)}{\Sigma(R; T_{H_2O}, P_{H_2O})}},
\end{equation}

\noindent where $X_i$ and $X_{H_2O}$ are the mass mixing ratios of the volatile $i$ and H$_2$O with respect to H$_2$ in the Solar Nebula, respectively. $\Sigma(R; T_i, P_i)$ and $\Sigma(R; T_{H_2O},
P_{H_2O})$ are the surface density of the nebula at a distance $R$ from the Sun at the epoch of hydratation or clathration of the species $i$, and at the epoch of condensation of water, respectively. From ${\it m_i}$, it is possible to determine the mass fraction of species $i$ with respect to all the other volatile species taking part to the formation of an icy solid:

\begin{equation}
{M_i = \frac{m_i}{\displaystyle \sum_{j=1,n} m_j}},
\end{equation}

\noindent with $\displaystyle \sum_{i=1,n} M_i = 1$.\\

It is important to mention that, after having performed calculations using a wide range of values from 10$^{-3}$ to 10$^{-1}$ for the viscosity parameter $\alpha$ of our turbulent model, we found that the volatile trapping conditions (temperature and pressure) remain almost constant at a given heliocentric distance in the Solar Nebula, whatever the adopted value of this parameter. In addition, Alibert et al. (2005a) demonstrated that the composition of icy planetesimals remains similar whatever their formation distance within the same disk, as long as a homogeneous gas-phase is postulated within the nebula. {\it These statements imply that, whatever the input parameters adopted when modelling the disk, and regardless of the formation location considered for icy planetesimals at distances beyond 5 AU, their composition (in wt\%) remains almost constant, provided that the gas-phase abundances are homogeneous in the nebula.}\\

\begin{figure}
\resizebox{\hsize}{!}{\includegraphics{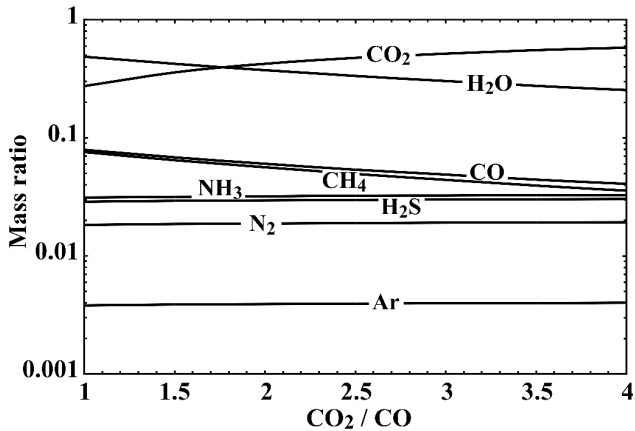}} \caption{Plot showing the composition of the ices (wt\%) incorporated in planetesimals produced in the Solar Nebula as a function of the adopted CO$_2$:CO ratio in the initial vapor phase. The composition is expressed \change{as a} mass ratio (mass of ice ${\it i}$ to the global mass of ices). The abundances of the considered elements are solar with CO:CH$_4$~=~1:1 and N$_2$:NH$_3$~=~1:1 in the vapor phase. Note that the mass fractions of
Kr and Xe are too low to be represented given the scale adopted in the figure.}
\label{comp}
\end{figure}

Figure \ref{comp} shows the variation of the composition of icy planetesimals formed in the outer Solar Nebula as a function of the CO$_2$:CO ratio (shown for values between 1:1 and 4:1), with CO:CH$_4$ = 1:1 and N$_2$:NH$_3$ = 1:1 in the gas-phase. The composition of ices given in this figure is then valid for solids formed at any distance within the outer Solar Nebula, provided that the gas-phase is homogeneous and the disk is initially warm enough at that location to vaporize the ices falling in from the ISM. It can be seen that water remains the most abundant ice in mass, provided that the CO$_2$:CO gas-phase ratio in the nebula is between 1 and $\sim$1.7. Interestingly enough, whatever the assumed CO$_2$:CO ratio, CO$_2$ remains the main carbon species trapped within planetesimals. We also note that the relative amounts of H$_2$O and of other carbon species decrease as the CO$_2$:CO ratio
increases. In contrast, the mass fractions of NH$_3$, N$_2$, H$_2$S, and of the noble gases are only weakly influenced by the variation of the CO$_2$:CO gas-phase ratio. Table \ref{fractions} gives the mass ratios of the ices for the two extreme cases CO$_2$:CO = 1:1 and 4:1.

\begin{table}
\caption[]{Ratio of the mass of ice ${\it i}$ to the global mass of ices (wt\%) in planetesimals formed in the outer Solar Nebula, calculated for CO$_2$:CO = 1:1 and 4:1 in the vapor phase. Both ratios are calculated with CO:CH$_4$~=~1:1 and N$_2$:NH$_3$~=~1:1 in the gas-phase of the nebula.}
\begin{center}
\begin{tabular}{lcc}
\hline
\hline
\noalign{\smallskip}
Species &  CO$_2$:CO = 1:1 & CO$_2$:CO = 4:1  \\
\noalign{\smallskip}
\hline
\noalign{\smallskip}
H$_2$O		&	48.8		&	25.5		 	\\
CO$_2$ 		&	27.5		&	58.4		 	\\
CO   			&    	7.9  		&   	4.2		 	\\
CH$_4$   		&  	7.6	        	&	3.7		 	\\
NH$_3$  		&    	\multicolumn{2}{c}{$\sim 3.3$}	 \\
H$_2$S	   	&   	\multicolumn{2}{c}{$\sim 2.9$}	 \\
N$_2$   		&    	\multicolumn{2}{c}{$\sim 1.9$}	\\
Ar 			&   	\multicolumn{2}{c}{$\ll 0.1$}  	\\
Kr 			&    	\multicolumn{2}{c}{$\ll 0.1$}  	\\
Xe  			&   	\multicolumn{2}{c}{$\ll 0.1$}  	\\	
\hline
\end{tabular}
\end{center}
\label{fractions}
\end{table}

\section{Delivery of icy particles to the Main Belt}

Particles in protoplanetary disks undergo orbital decay due to the
effects of gas drag (Weidenschilling 1977). The efficiency of this
process is particularly dependent on the gas density inside the disk,
and on the size and density of the particles. Cyr et al. (1998)
showed that icy particles produced in the outer Solar system may have
been incorporated in bodies formed in the inner regions. Indeed,
taking into account gas drag, sedimentation and sublimation, these
authors calculated that particles originating from $\sim 5$ AU may
migrate to heliocentric distances of the order of 3 AU due to gas
drag, before subliming as a consequence of the higher temperature and
pressure domains encountered in the inner regions. However, it is
important to note that Supulver \& Lin (2000), Cuzzi et al. (2005) and
Ciesla \& Cuzzi (2006) were unable to reproduce these
results. More recently, MA05 examined the effect of gas drag
on the migration of icy particles using a time dependent model of the
solar nebula\footnote{The work of Cyr et al. (1998) is based on the
static Solar Nebula models elaborated by Cassen (1994). In
each of these models, the position of the condensation front of water
(the ``snow-line'') remains at a constant heliocentric distance. In
contrast, evolutionary models cool with time and allow the snow-line
to migrate closer to the Sun.}. They demonstrated that, after
the Solar system has experienced a few Myr of cooling, it is possible for icy particles from the outer Solar system
to drift inward to the present location of Ceres ($\sim 2.7$ AU). In
such an evolved system, because the snow-line has already
passed through the Main Belt, the volatiles contained in the icy
particles can avoid sublimation during their migration.

In this paper, we extend the calculations originally performed by
MA05, in order to take into account the possibility of volatile
capture directly in the form of pure ice, in case there is too little
water to trap all volatiles as hydrates and clathrate
hydrates. Moreover, for the sake of consistency, we use the disk model
that allows the formation of Jupiter and Saturn, leading to planets
with internal structures consistent with observations (see Alibert et
al. 2005b).

\begin{figure}
\centering
\resizebox{\hsize}{!}{\includegraphics[angle=0]{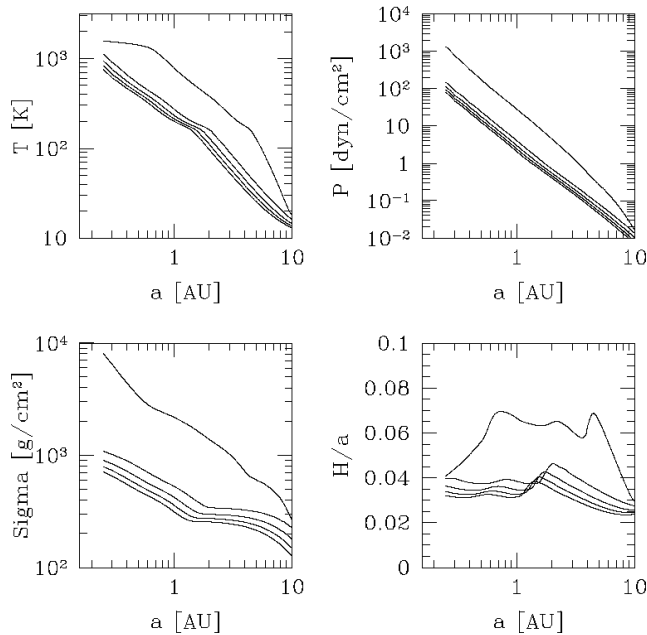}}
\caption{Thermodynamic conditions in our disk model. Top
left: midplane temperature profile, top right: midplane pressure
profile, bottom left: surface density profile, bottom right: aspect ratio
{\it H}/{\it a} profile ({\it H} is the semi thickness and {\it a} the heliocentric distance). The epochs shown in the 
four panels are, from top to bottom, 10$^4$ yr, 0.5, 1, 1.5 and 2 Myr.}
\label{disk_structure}
\end{figure}

The numerical procedure used to calculate the structure and evolution
of the disk are entirely detailed in Alibert et al. (2005c), and in
this work we only give the main points. More details can be found in
the afore-mentioned paper. The surface density $\Sigma$ in the disk is
calculated by solving the diffusion equation:

\begin{equation}
 {d \Sigma \over d t} = {3 \over r} {\partial \over \partial r } \left[
 r^{1/2} {\partial \over \partial r}
 ( \tilde{\nu} \Sigma r^{1/2})  \right]  + \dot{\Sigma}_w(r),
 \end{equation}
 
\noindent where $\tilde{\nu}$ is the mean viscosity (integrated along
the $z$ axis), and $\dot{\Sigma}_w(r)$ is the photoevaporation term,
taken as in Veras \& Armitage (2004). The viscosity is calculated in
the framework of the $\alpha$-formalism (Shakura \& Sunyaev 1973)
after first solving for the vertical structure of the disk (see
Alibert et al. 2005c). This calculation also gives us the evolution of
the thermodynamic conditions inside the disk, as a function of time
and heliocentric distance. These conditions are used, together with
information on the composition of the nebula's gas, in order to derive
the amount of volatile species trapped inside icy planetesimals. For
the model we use here, the $\alpha$ parameter is equal to $2 \times
10^{-3}$, and the total evaporation rate is of the order of $10^{-8}
\Msun$/year. At the beginning of the calculation, the gas surface
density is given by a power law, $\Sigma \propto r^{-3/2}$, normalized
to have $\Sigma = 600$g/cm$^2$ at the current day position of
Jupiter. As stated above, this disk model was used in Alibert et
al. (2005b) in order to calculate formation models of Jupiter and
Saturn. The resulting thermodynamic conditions are plotted in
Fig. \ref{disk_structure} for early epochs during the
evolution of the Solar Nebula.

\begin{figure}
\centering
\resizebox{\hsize}{!}{\includegraphics[angle=0]{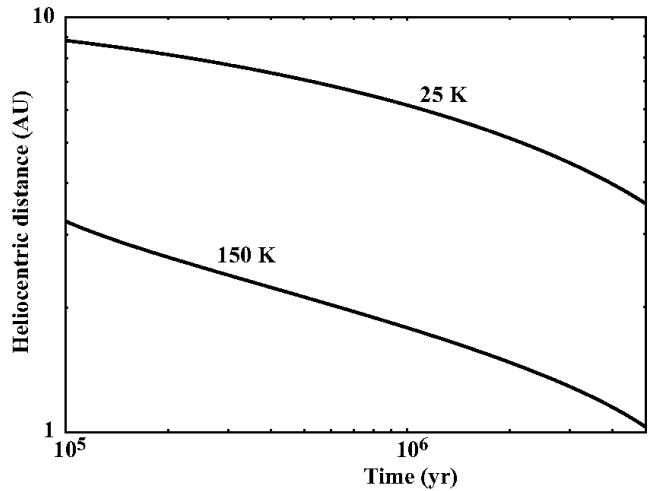}}
\caption{Locations of the 25K- and 150K-lines as a function of time for our model of the primitive nebula.}
\label{icelines}
\end{figure}

Figure \ref{icelines} shows the location of the snow-line\footnote{The
sublimation temperature of water ice is taken as 150K, and does not
depend on pressure in our calculations.} and the 25K-line\footnote{At
temperatures lower than $\sim$25 K, all the main volatile species are
presumed to be trapped as clathrate hydrates or in the form of pure
condensates in the Solar Nebula (e.g. Fig. 1).}  as a function of
time. In our calculations, we assume that when a particle reaches the
snow-line (or the 25K-line), it loses all its volatiles (or loses
all the volatile species whose condensation temperatures are lower
than or equal to 25 K). This is of course a simplification, and more
accurate calculations should take into account the progressive heat
diffusion, and the resulting sublimation, within icy particles
entering the warmer regions of the early Solar Nebula. That such low temperatures are reached at small heliocentric
distances in the Solar Nebula can be surprising, given that
the equilibrium radiation temperatures are of $\sim$160 K at 3
AU. However, our calculations are justified by the fact that,
during the lifetime of the nebula, the Sun's radiation is
significantly attenuated, even at distances of a few AU, due
to Rayleigh scattering from molecular hydrogen and dust opacity, in
particular if no inner gap is postulated in the disk (Mousis et
al. 2007). In these conditions, light becomes extinguished close to
the star as a result of the high gas density, while the outer regions
play little role in the extinction. For temperatures below 1500~K, the
dominant dimming effect in the nebula at wavelengths shorter than a
few $\mu$m is Rayleigh scattering from molecular hydrogen (Mayer \&
Duschl 2005). This condition is fulfilled for the entire nebula after
$10^5$~yr, with the temperature beyond 0.6~AU already falling
below 1000~K at this early stage. Moreover, the mass absorption
coefficient has been estimated to be $\sigma_m = 5 \times 10^{-4} \rm
(cm^2/g)$ in the Solar Nebula (Mousis et al. 2007). Integrating along
the mid-plane radial axis of the model from 0.25 AU (the inner edge of
the disk), the optical depth at 3 AU is initially about $\tau =
18$. It decreases over time to $\tau = 5.3$ after $\sim$5 Myr of the
disk's evolution. As a result, only $\sim$0.5\% of the star's
radiation will be available at this distance within the disk
at this epoch. We can therefore assume that the disk remains
optically thick to Solar radiation during at least the first
5 Myr of its evolution. Note also that it has been recently showed 
that disks with puffed up inner edge could shield part of their outer 
edge from star light, in agreement with the structure and the overall 
spectral energy distribution (SED) of protoplanetary disks around 
Herbig Ae/Be stars (Dullemond \& Dominik 2004). This is the case for 
our solar nebula model which is self-shadowed at heliocentric distances 
higher than $\sim$2--3 AU (see semi thickness over heliocentric distance 
profile in Fig. \ref{disk_structure}). Since the Rayleigh scattering, 
as well as the strong opacity of grains impede the light to go through 
the upper layers and illuminate the outer parts of the disk, the cooling 
of the nebula down to low temperatures should follow, even in the 
formation zone of asteroids. 

\begin{figure}
\centering
\resizebox{\hsize}{!}{\includegraphics[angle=0]{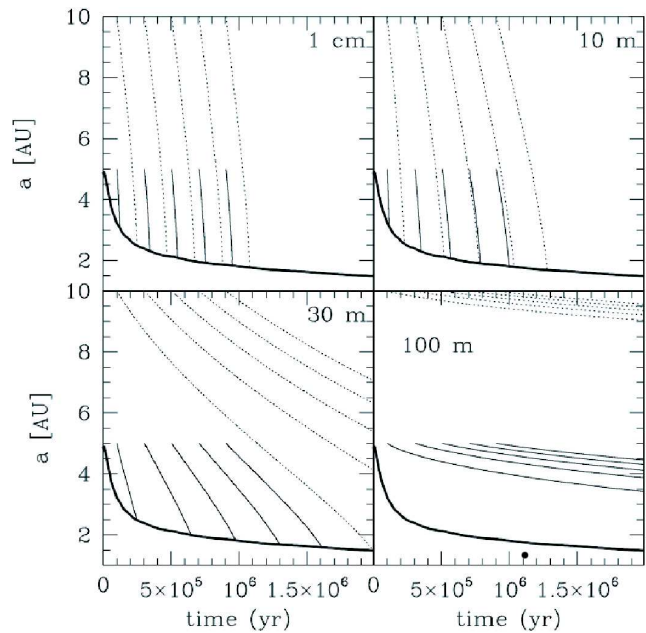}}
\caption{Trajectories of solid particles in the primitive nebula, for different starting location and starting epochs. The initial location of particles is either 5 AU or 10 AU, and the drift starts during the first Myr of the Solar Nebula evolution. The trajectories are drawn until particles cross the iceline (heavy solid line), from which they are assumed to vaporize. The size of particles is given in each panel, and their density is equal to 1 g cm$^{-3}$.}
\label{radialdrift}
\end{figure}

In our calculations, we have considered particles of various
sizes, ranging from 1 cm to 100 m. We assume that they are formed at
different locations in the nebula, and at different epochs, in a
similar way to MA05. Figure \ref{radialdrift} shows the trajectories of particles of size 1 cm, 10 m, 30 m and 100
m. For particles ranging between a few centimetres and a few
metres in size, gas drag is so strong that the trajectories would be
represented as vertical lines. The trajectories of the particles are
stopped when they cross the iceline. Figure \ref{radialdrift} shows
that, due to gas drag, particles with the sizes considered here can
drift to the present day location of the Main Belt. The innermost
location of particles that have not suffered any sublimation of ice is
therefore given by the location of the snow-line at a given
time. Similarly, the 25K-line gives, as a function of time, the
innermost location of particles that have not lost any of their
volatiles. The situation is more complex for particles massive enough
to significantly decouple from the gas, which therefore experience
minimal, if any, orbital decay (see last panel - 100 m size objects -
in Fig. \ref{radialdrift}). Icy material can thus be
transported by such medium size particles. If they cross the
25K-line, they would be expected to lose some fraction of
their volatile content. However, particles formed even later, after a few Myr of
the evolution of the Solar Nebula, but before it becomes
optically thin, do not encounter temperatures higher than 25K. They
can therefore preserve all of their volatile species from vaporization
until their capture by forming asteroids. 

On the other hand, even if icy particles may have drifted to heliocentric distances low 
enough to take part to the formation of the whole Main Belt without losing a non negligible fraction of their volatiles, the current distribution of ices that actually exists in 
asteroids has probably been significantly altered during and/or after their formation. Indeed, 
more heat was generated during the accretion of asteroids located at low heliocentric distances, as impacts and collision velocities were greater. As a result, vaporization has occurred, thus changing the composition of ices incorporated in inner belt asteroids. Moreover, the equilibrium
radiation temperatures reached on the surfaces of Main Belt asteroids
after the dissipation of the nebula can be too high for the long-term
stability of water ice, particularly in the case of inner belt
asteroids (Jewitt et al. 2007). Indeed, even if potential internal
heating from the decay of Al$^{26}$ and other isotopes is not
considered, radiation equilibrium may have led to the vaporisation of
the ice content of the nearer asteroids (semi-major axes of $\sim$2
AU), and melted the ice of mid-range asteroids situated at $\sim$3
AU. However, it will probably not have affected the ice in the
interiors of asteroids located further out. Inner and outer asteroids
would therefore display no detectable hydratation features either
because the ice was vaporized and dissipated, or because the ice never
melted and thus did not react with other minerals that would enable
its detection (Cyr et al. 1998). On the other hand, mid-range
asteroids could be expected to have undergone sufficient melting that
chemical alteration of silicates would occur and be detectable (Cyr et
al. 1998; Jewitt et al. 2007). Water ice could also be present
temporarily on the surface of large hydrated bodies like Ceres, while
migrating outwards from sub-surface layers or mantle of the asteroids,
and before it sublimates into space (Fanale \& Salvail, 1989).

These considerations imply that the future detection and identification of the volatile phases in asteroids could give some constraints on the thermodynamic conditions that were present within the Solar Nebula during their accretion, as well as during the inwards migration of icy planetesimals. For example, assuming a cold accretion for asteroids located in the outer belt and that the only detected volatile species on the surfaces of some of their members are CO$_2$; H$_2$S, NH$_3$ and H$_2$O, one may deduce
that the temperature of the nebula on the migration pathway of icy planetesimals was between $\sim$60 and 80 K (e.g. Table \ref{cond}). However, we note that alteration of the volatile phases
in asteroids may occur after their formation, as a result of catalytic reactions in their interiors. Indeed, this mechanism has been proposed to explain the current composition of the plumes released by the Saturn's moon Enceladus (Matson et al. 2007) and could occur at early epochs following the accretion of asteroids due to internal heating caused by the decay of $^{26}$Al. Hence, the influence of this chemistry within the interior of asteroids may also play a role in the composition of the volatile phases that potentially exist in some members of the Main Belt.

\section{Implications for the porosities of asteroids}

This Section is devoted to a discussion of the determination of the taxonomical types of the asteroids, and their densities. It appears that some of the results in this area, due to their low accuracy, provide poor constraints on the internal structures of the asteroids in question, and it is therefore difficult to infer their composition from these measurements. This is particularly true for the M type asteroids which often show a puzzlingly high albedo, and a measured bulk-density much lower than the densities of their supposed analogue metallic meteorites. This implies that our current knowledge of the physical properties of asteroids is not in contradiction with the idea that some of them may have preserved substantial amounts of the ices accreted during their formation. We then discuss the potential influence that the incorporation of ices in asteroids may have on their resulting densities and porosities.

\subsection{Uncertainties in the taxonomical class and densities of asteroids}

The computation of porosity necessitates the knowledge of both the bulk-density of an asteroid and the bulk-density of its constituent material. With the exception of the few 'giant' objects or dwarf planets (Ceres, Pallas and Vesta), the bulk-density of asteroids is generally smaller than that which would be expected from the densities of supposed meteorite analogues. The relationship between meteorite analogues and sample densities is, however, not straightforward and can, in some cases, be misleading. This is particularly important for the objects classified as M, P or E-type asteroids in the Tholen classification scheme.

 Recently, a classification based on spectral morphology has been proposed by Bus et al. (2002), where these asteroids are related to class X and sub-classes therein. Asteroids within these classes generally display a linear trend in their reflectance curve, and often also have high albedo. In the case of radar observations, some M-type asteroids also showed particularly strong echo signatures, suggesting
they are essentially made of metal, just like the M-type meteorites. However, this might not be always the case, and some X-type asteroids could also be closer to primordial D- or P-type objects, and hence have lower densities. In any case, their connection to meteorites is either unknown or unclear (Clark et
al. 2003; Hardersen et al. 2005) and has been under debate during the last decade. M-type asteroids were generally thought to be the remnants of the metallic cores of differentiated asteroids or planetesimals that would have lost their silicate mantle, possibly after a disruptive catastrophic collision. There is still debate about the composition and origin of M- or X-type asteroids (Busarev 1998; Rivkin et al. 2000). Following Lupishko \& Belskaya (1990), a purely metalic composition is hard to believe, and E-type chondrites or stony-iron meteorites could be better analogues. Nowadays it appears likely that these classes comprise an assemblage of several possibly unrelated surface types with a featureless spectrum, not all of which are metallic. Indeed, some asteroids previously classified as M-type have shown spectra with the $3\,\mu$m signature associated with aqueous alteration which clearly makes this metallic-core hypothesis less viable. Similarly, 21 Lutetia has been classified as an M-type asteroid since it's albedo is relatively high, but recent spectroscopic observations suggest it is actually a
classical C-type body (Lazzarin et al. 2004).

On the other hand, the bulk-density of an asteroid is derived from measurements of its mass and size. Mass determinations obtained as a result of the dynamical perturbations experienced during close encounters are less accurate, particularly when they are based on a single target perturbation (Hilton 2002). One can find large discrepancies when comparing the values obtained by this technique to the one derived from the analysis of a moonlet's orbit. For example, the masses of 22 Kalliope and of 87 Sylvia differ by a factor 2 between Marchis et al. (2003) and Kochetova (2004), or between the results of Marchis et al. (2005) and Ivantsov (2007), where the latter values should theoretically be the less accurate ones. There are also large discrepancies between different results obtained from the same technique but involving different observations: depending on the authors, the mass of Psyche ranges from $(0.68\pm0.14) \times 10^{-11}$ to $(0.87\pm0.26) \times 10^{-11}$ or $(1.49\pm0.31) \times 10^{-11} \Msun$ (Lupishko 2006) or even up to $(3.38\pm0.28) \times 10^{-11}$ or $(4.0\pm1.4) \times 10^{-11}$ \Msun (Kuzmanoski \& Kova{\v c}evi{\'c} 2002; Ivantsov 2007), all values that are generally outside the usual statistical 3-sigma margin. Finally, an error in the size of the body (mainly resulting from the determination of the object's albedo) will introduce a large error in the determination of the bulk-density. We find that the revised density of Psyche from the polarimetric albedo of Lupishko (2006), together with the largest value for the mass of Kuzmanoski \& Kova{\v c}evi{\'c} (2002), would then reach the unrealistic value of $\rho=13.2$g/cm$^3$. Nonetheless it is stressed that the matching between the IRAS-based diameters and the sizes measured from resolved asteroids in the Main Belt is generally good (Cellino et al. 2003; Marchis et al. 2006), in contradiction with the latter disagreement of Lupishko (2006) for this particular M-type asteroid.

It therefore seems that it is difficult to obtain meaningful statistics using the taxonomic M-type class because diversity is still present (the situation is barely improved when one considers Bus et al.Õs X-class). It is therefore clear that an average value for the density of the M-type asteroids (e.g. Krasinsky et al. 2002) can be misleading. It would also be difficult to derive knowledge of the internal structure of these bodies since the bulk densities quoted in the literature can often be in error.

\subsection{Influence of the volatile content in asteroids on their bulk densities}

Considering the delivery scenario detailed in section three, the
asteroids would have been accreted from a mix of icy solids that
formed initially in the outer Solar Nebula, and that preserved their
volatile content during their inward migration, with heavier and
possibly differentiated material produced at smaller heliocentric
distances. Such a scenario would not, however, be applicable to the
gravitationally re-accumulated aggregates that result from a
catastrophic collision (such as rubble-pile asteroids, and possibly
binary objects), because their formative collisions would
happen at a later stage of the evolution of the solar
system. Nevertheless, ices could still be present in rubble-pile
asteroids even after \change{exposure} to solar irradiation.

The time scale for re-accumulation scales as
$(G\,\rho)^{-1/2}$, where $G$ is the gravitational constant and $\rho$
the density of the body, so that the dynamical time scale for the
re-accumulation of the aggregates is -- depending on their actual
velocity dispersion -- of the order of several days (Michel et al
2001, 2004; Durda et al. 2004; Nesvorn{\'y} et al. 2006). This
corresponds roughly to a few revolutions of the debris (the
remaining material would be ejected). At times larger than several
weeks, debris are no longer accumulated, but rather dispersed
into a dynamical family. The time scale for re-accumulation after a
catastrophic disruption of an asteroid is thus of the order of
weeks. On the other hand, the sublimation rate experienced by
the debris depends on several conditions: temperature (heliocentric
distance), and the presence and thickness of soil and regolith
(Chevrier et al. 2007). The sublimation rate for a planar surface of
pure ice is proportional to $P_s.(M/2\pi R T)^{-1/2}$ in a vacuum
(e.g. Novikov \& Vagner 1969, Patashnik \& Rupprecht 1975; Farmer
1976), where $P_s$ is the saturation vapour pressure at $T$, $M$ the
molecular weight, $R$ the universal gas constant, and $T$ the
temperature.

Depending on the author, the coefficient factor used to
calculate the sublimation rate from this proportionality incorporates
different effects (such as the rugosity, presence of other gas, soil
shell burying the ice, etc) and can vary within two order of
magnitude. The evaporation rate is, however, of the order of one
metre/year at most (Andreas 2007; Hsieh \& Jewitt 2006). At 160 K,
which is a typical temperature for a moderate albedo body in the
Main Belt, the recession rate of water ice can be as low as a few
mm/year (Hsieh et al. 2004). The sublimation life-time for kilometre
sized bodies is therefore of the order of a few thousand years.

The dynamical time for re-accumulation is therefore much
shorter than that for the sublimation of all the ice in collisional
debris. As a consequence, one can neglect for our purpose the
evaporation of ices during the post-impact re-accumulation
phase.  We argue that water and other volatiles are not easily
revealed, and that such hidden ices will have a significant influence
on the internal structure and porosity of the asteroids. Such a
scenario was suggested by Veverka et al. (1997) for the asteroid 253
Mathilde, but rejected by the authors in the absence of any alteration
feature in the spectra. Similarly, Wilson et al. (1999) consider the
case where all volatiles must have been completely lost, thus
increasing the porosity, but again only for the altered bodies.
In general, the presence of an unrevealed icy fraction will
reduce the expected bulk-density (the body's average mass-to-volume
ratio), regardless of its mass distribution, as would do the presence of
voids and cracks, or any other light material such as a
regolith. Thus, a low measured density could be explained by voids or
cracks in a rubble-pile structure, or a fractured body, or by the
presence of light ices, or both, since these hypotheses are
not mutually exclusive. In the following discussion, we address the
bias introduced by the presence of ices in a simple computation of the
macro-porosity of an object.

The macro-porosity of an asteroid is calculated by the determination of the fraction of empty space within the total volume, which translates to $1-{\rho_{b} /\rho}$ where $\rho_{b}$ and $\rho$ are the measured bulk-density of the asteroid and the assumed sample bulk-density, respectively. Following Britt et al. (2002), the (uncompressed) sample density is derived through knowledge of the asteroid's taxonomic type, using values measured in meteoritic analogues, and taking into account microporosities in the meteoritic grain. This macro-porosity can now be computed by taking into account a possible volume fraction of ices $f_{i}$ or silicates $f_{s}=V_{s}/(V_{s}+V_{i})$ (where $V_{s}$ and $V_{i}$ are the volume of silicates and ices, respectively), so that $f_{s}+f_{i}=1$. This corresponds to an effective sample density $\bar\rho_{s}=f_{s}\,(\rho_{s}-\rho_{i}) + \rho_{i}$, and the bulk-porosity is now given by:

\begin{equation}
  \psi = 1 - {\rho_{b}\over f_{i}\,(\rho_{i}-\rho_{s}) + \rho_{s}}
  \label{E:poro}
\end{equation}

\noindent where one finds the limiting values of porosity for
$f_{i}=0$ and $f_{i}=1$.

\begin{figure}
\centering
\resizebox{\hsize}{!}{\includegraphics{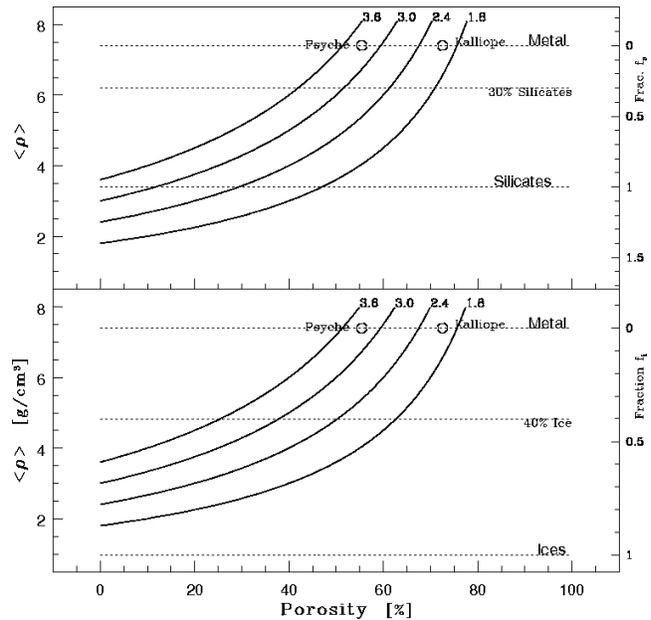}}
\caption{The porosity of asteroids considering a possible
volume-fraction of ices ($f_i$) or silicates ($f_s$) of density
$\rho=0.97$ or 3.4, respectively. The ordinate is the average sample
density $\langle\rho\rangle=f_i\,(\rho_i-\rho)+\rho$. The different
curves correspond to various values of measured bulk-densities
($\rho_b$) of asteroids. Upper panel: porosity for M-type asteroids
with both metal ($\rho = 7.4$) and \change{silicaceous}
components. Bottom panel: porosity for M-type asteroids with both
metal and ices.}
\label{F:poro}
\end{figure}

As can be seen in Fig.~\ref{F:poro}, the bulk porosity of an object
with a significant icy component can remain relatively low, in the
fragmented regime of Britt et al. (2002).  The corresponding mass
fraction is $ X_{i} = \rho_{i}/\rho_{s} \cdot f_{i}/(1-f_{i}) \approx
(3\,(1/f_{i}-1))^{-1}$ for typical values of ice and silicates. Thus
the value of $X_{i}=17-27\,$\% for the water content by mass
determined by McCord \& Sotin (2005) in the case of Ceres translates
to $f_{i}\approx40\,$\% in volume. Note that the values
shown in the figure are not strictly restricted to water
ice. Indeed, the densities of the considered hydrates, clathrate
hydrates or pure ices are very close to that of water ice ($\sim$0.9
g/cm$^3$). Nevertheless, if a large fraction of ice must be present in
a given body, it is most likely that they would consist of
H$_2$O and CO$_2$, as shown in Table~\ref{fractions}.

\begin{figure}
\centering
\resizebox{\hsize}{!}{\includegraphics{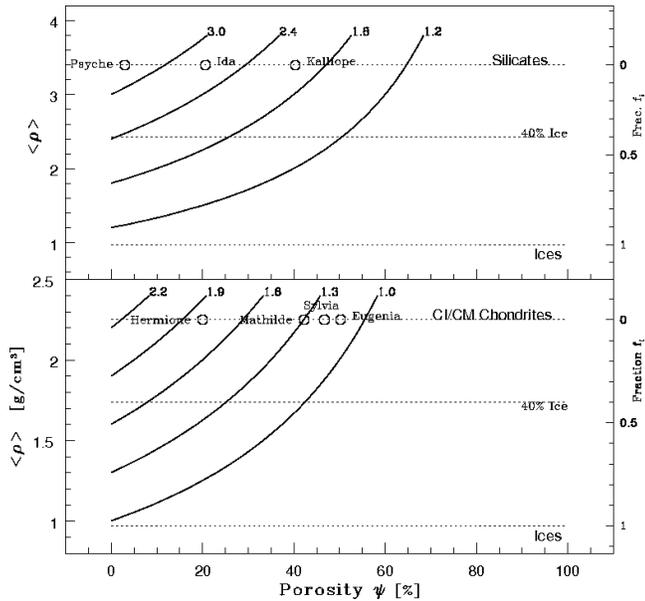}}
\caption{Same as Fig. \ref{F:poro} with the volume-fraction of ices
($fi$) and silicates ($\rho = 3.4$) or the lightest chondrites ($\rho
= 2.2$). Upper panel: porosity for S-type asteroids with some icy
fraction (right legend). Bottom panel: porosity of C-type asteroids
with ices. }
\label{F:poroch}
\end{figure}

Assuming a mass fraction of silicate or ices together with metallic
components would lead to a reduction of the mean sample density
$\langle\rho\rangle$, and hence the computed bulk-porosity. As shown
in Fig.~\ref{F:poro}, the porosity of M-type asteroids 22 Kalliope and
16 Psyche (both X-type in the Bus et al. classification) with the
adopted updated bulk-densities of $\rho=2.03\,$g/cm$^3$ (Marchis et
al. 2003) and $\rho=3.3\,$g/cm$^3$ (Lupishko 2006), respectively,
could be reduced by considering a substantial volume-fraction of
either silicates (olivine, pyroxene) or ices (water, ...). In the case
of Kalliope, however, only a large mass fraction of silicates or ices
could reduce the porosity to a realistic packing for a supposedly
post-collisional gravitationally bound granular material (i.e. less
than approximately 50\%). Psyche is a controversial object because its
radar echo is high, which makes a metallic surface plausible. Keeping
in mind that the error bars are so large that its density could be in
the range $[1.7-13.2]$, it is clear that the lowest values are
feasible only if a significant fraction of lighter material is
introduced. Moreover, in this latter case, there is no reason to
expect a porosity larger than 20-30\%, i.e. larger than that assumed
for fractured objects. Furthermore, considering that these bodies'
main constituent is of a lower density (closer to those of silicates
or stony-iron meteorites $\sim 3.3-4.5$), one sees that these could
still contain a fraction of the order of 30\% ice.

Similarly, one sees that for the low density objects like Eugenia, Mathilde and Sylvia and the C-type asteroids in general, introduction of a significant mass fraction of ices would change the computed porosity of these bodies. Note that a composition of 30\% water ice and 70\% CI/CM-type chondrites in the Cb-type asteroid 253 Mathilde would imply an average porosity of approximately 30\% for both constituents, a value which seems plausible; at least for the porosity of the ice. Pushing this exercise further, one can also derive either the density of the other constituents, by assuming a constant ice content and a constant porosity, or inversely, derive the mass- or volume-fraction of ices at constant macro-porosity and a given heavy element (chondrite, silicate, stony-iron, etc.)  density. For instance, the C-type asteroids Mathilde, Eugenia and the X/P-type Sylvia could all have the same density within their heavy constituents, and a porosity of 40\%. This would then require them to contain approximately 5, 15 and 45\% water-ice, respectively.

\begin{figure}
\centering
\resizebox{\hsize}{!}{\includegraphics{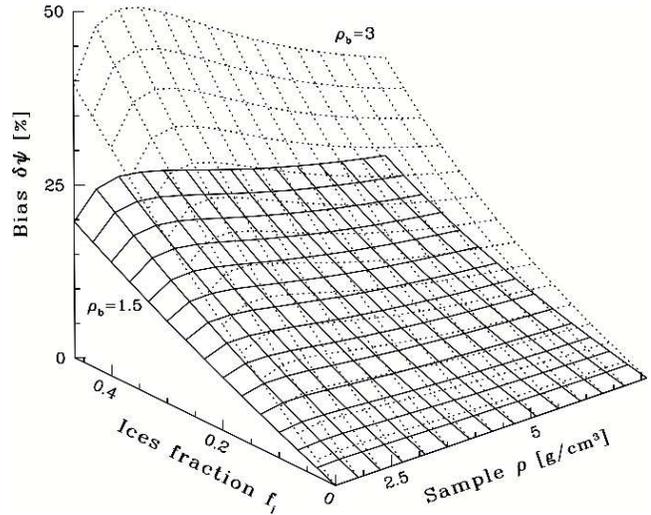}}
\caption{The bias introduced in the computation of the porosity ($\psi$) by neglecting a possible presence of ices (Eq. 4). Its resulting value is given as a function of the actual volume fraction of ices ($f_{i}$) and the sample density ($\rho_{b}$) of the body. It is plotted here for two cases of $\rho_{b}$ (solid and dotted grid).}
\label{F:biais}
\end{figure}

In conclusion, water and other volatiles could have remained buried inside the Main Belt asteroids since the formation of the solar System without being easily detected by remote spectroscopic observations. The presence of such ices will reduce the value of the bulk density of the body that would be expected from its taxonomic type, and, moreover -- given the small amounts of compaction inside asteroids -- it will additionally decrease the asteroid's macro-porosity (Durham et al. 2005). There would be no a priori
preference of this scenario for any asteroidal taxonomic type (hydrated or not), or collisional evolution (given short re-accumulation time-scales). On the other hand the size and mass of the asteroid may play an important role, which is not addressed here. Neglecting the possible presence of such ices can thus increase the error biases on the computation of the porosity, leading to it being systematically overestimated. For small amounts of ice ($f_i \sim0.2$, see Fig.~\ref{F:biais}), the error in the porosity is
$\delta\psi\sim f_{i}\,\rho_{b}\,(\rho-\rho_{i})/\, [\rho^2\,(1+f_{i}\,(\rho_{i}-\rho)/\rho)] $ which yields a bias in the calculated porosity of at least 5 to 15 percent depending on the actual bulk-density of the asteroid ($\rho_{b}$), its supposed sample density ($\rho$), and obviously on the actual volatile component ($f_{i}$). In addition, the precision of asteroids bulk densities measurements is increasing with better knowledge of masses and current high resolution angular measurements for the size (not to mention space probes). As a consequence, the importance of the error or bias discussed here should become more and more important.

\section{Summary and discussion}

In order to explain the presence of hydratation and cometary features in the Main Belt, we have proposed that asteroids incorporated during their formation icy particles formed in the outer Solar Nebula. We have then calculated the composition of the ices trapped in these planetesimals formed beyond a heliocentric distance of 5 AU in the nebula, in a manner consistent
with the formation of Jupiter, by assuming that the gas-phase abundances of all elements, in particular that of oxygen, are solar. As a result, we have found that the ices being formed in the outer Solar Nebula are composed of a mix of clathrate hydrates, hydrates formed above 50 K, and pure condensates produced at temperatures between $\sim$20 K and $\sim$50 K. We have noted that, whatever the input parameters adopted in the modelling of the disk, or the formation location considered for icy planetesimals at heliocentric distances beyond 5 AU, their composition remains almost constant, provided that the gas-phase abundances are homogeneous in the nebula. We have argued in this work that gas-drag is responsible for the inward drift of icy particles formed in the outer nebula towards the forming Main Belt. To support this hypothesis, we have showed that, at some epochs of the disk's
evolution, some particles produced in the outer nebula may drift to the current position of the Main Belt without encountering temperature and pressure conditions high enough to vaporize the ices they contain.

The current distribution of ices potentially existing in asteroids has probably been deeply altered after their formation. The effect of solar insolation may have vaporized the ice within nearer asteroids (semi-major axes of $\sim$2 AU), melted the ice of mid-range asteroids situated at $\sim$3 AU, but should not have affected the ice in asteroids located at greater heliocentric distances. Inner and outer asteroids would therefore display no detectable hydratation features, either because the ice was vaporized and dissipated, or because the ice never melted and thus did not react with the surface minerals to a sufficient extent as to allow detection (Cyr et al. 1998). In this context, we have proposed that, from the detection and identification of initially buried ices revealed by recent impacts on the surfaces of asteroids, it could be possible to infer the thermodynamic conditions that occurred within the Solar Nebula during the accretion of these bodies,
as well as during the inwards migration of the icy planetesimals which they incorporated. However, this statement requires that either no parent body processing or modification took place during and after the formation of asteroids. For example, we have noted that subsequent alteration of the volatile phases in asteroids may occur due to catalytic reactions in their interiors. 

We have also investigated the potential influence that the incorporation of ices in asteroids may have on their porosities and densities. In particular, we have showed that the presence of ices can considerably reduce the value of the bulk density of the body, and consequently its macro-porosity, that would be expected from a given taxonomic type.

That volatiles were delivered to areas within the ice line is clearly beyond doubt. In addition to the gas-drag mechanism described in this work, it is also likely that a significant amount of volatile material was
dynamically driven inwards in the latter stages of planet formation. We still see the tail of this dynamical, chaotic volatile movement today -- the comets we observe passing through the inner solar system are the bearers of ices formed far beyond the snow line, and held in deep freeze since the early days. During the latter stages of planetary migration, the flux of such objects passing through the inner solar system, and hence encountering the asteroids, was significantly higher. Of particular interest, when one considers veneers of volatile material near the surface of the asteroids, is the Late Heavy Bombardment. In the Nice model, (see e.g. Gomes et al, 2005), vast amounts of volatile-rich material is flung inwards from the outer solar system approximately 700 Million years after its birth. This event, caused by the resonant destabilisation of the outer solar system, would have coincided with a simultaneous stirring of the asteroid belt, leading to an impact flux upon the Earth containing approximately even proportions of asteroidal and cometary material. It is clear, though, that the Earth would not be the only object to encounter volatiles injected in this way, and the possibility of a late-veneer of ice arriving in the asteroid belt is surely something which must be acknowledged in future work. In addition to this aggressive and chaotic injection of material, there is also aÞ gentler mechanism by which volatiles can be driven inwards as a result of planetary migration. As planets migrate, material can be trapped in the locations of mean-motion resonances (MMR), which sweep in front of the planet through it's motion. Evidence of material being swept outwards in the resonances of Neptune is clear for all to see -- the Plutino family of objects are locked in the 2:3 MMR with the planet, and have an inclination distribution which can tell us a great deal about the distance over which the planet migrated, sweeping them along. Inward migration can have the same effect -- the interior resonances of a planet can collect material as it moves inwards, and sweep it along -- giving a mechanism by which volatile material can be eased inwards, with the migration of a giant such as Jupiter. Work such as Fogg \& Nelson (2006) has shown that such resonant forcing can operate with a resonable efficiency, even for significantly faster migration than expected in our solar system, and so the effects of this behaviour should not be ignored in future work.

In spite of the growing pool of evidence pointing towards the existence of water ice in the Main Belt, its detection on asteroids is a challenging observational problem. Large bodies such as Ceres are suspected to have retained a large amount of water since their formation, perhaps even including an internal liquid ocean, throughout the age of the solar system (McCord \& Sotin 2005). This could particularly be the case if this internal water was originally mixed with some ammonia, in agreement with our composition calculations in section \ref{icy}, which would have the effect of lowering the melting point of the water (ammonium bearing minerals have been suggested by King et al. (1992) as an alternate explanation for the origin of the 2.07 $\mu$m band seen in the spectrum of Ceres). Nevertheless, internal water can only be indirectly probed, either by measuring the hydrostatic shape of the object, as was done in the case of Ceres, or by inferring its density from its size and mass, when they are known or, more evidently, from outgasing activities. The case of Ceres is particularly interesting since, in spite of several possible pieces of evidence which support it being a highly hydrated body, the
only report of water detection on the dwarf planet is the observation of OH escaping from its northern pole\footnote{An OH atmosphere was indeed observed around Ceres after perihelion by A'ÕHearn and Feldman (1992) by performing IUE long exposure spectra, with column densities of the order of $10^{11}$~cm$^{-2}$.}, is still not confirmed. Nevertheless, this detection could be explained in the
context of the accumulation of ice during winter on the surface or within the subsurface layer, which would then dissipate during summer, when the surface temperature rises. Similar transient events have been suggested as possible mechanisms to trigger the geyser-like activity taking place near the south pole of Enceladus and reported by Cassini (Porco \& Team 2006).

It is interesting to note that, considering the gravity and the day-side temperature of Ceres, any outgassed atmosphere would be rapidly lost. The mean thermal velocity $v_0$ of H$_2$O, for instance, would be close to the escape velocity  ($v_{\infty}=516$~km~s$^{-1}$). Assuming a subsolar temperature of 215~K (Dotto et al. 2000), $v_0$ would vary between 450 to 350~km~s$^{-1}$ from the subsolar point to a zenith angle of 80$^{\circ}$. As a consequence, hydrodynamical escape would occur ($v_{esc}^{2}/v_{0}^{2} \le 2$). The photolysis of H$_2$O by solar EUV makes this atmospheric escape even more efficient by giving the photodissociation products OH and H some additional kinetic energy. Considering the short lifetime of H$_2$O at $\sim$3~AU ($<$9~days), and the fact that the mean thermal velocity of H atoms exceeds $v_{\infty}$, a tenuous atmosphere of OH is expected if water is outgassed by the asteroid at a sufficient rate. Due to the transcient nature of the atmosphere, the loss of water to space is limited by the flux of water from the interior to the surface. At low latitude, where ice is not stable, the continuous flux of water from the interior to space is too low to be detected.  Only an accumulation of water ice at high latitude before perihelion, followed by an outgassing of H$_2$O  associated with post-perihelion warming seems to result in an observable column density of OH. These results were found to be consistent with an earlier work done by Fanale and Salvail (1989), who estimated the mean loss rate of H$_2$O to be in the range 30-300~g~s$^{-1}$. Even if one assumes that the atmospheric loss observed by  AÕ'Hearn and Feldman (1992) occurs continuously at the same rate and at all latitudes (which is obviously wrong as this maximum loss requires high latitudes and post-perihelion conditions), the water loss remains below 4 kg~s$^{-1}$, which, integrated over 4.5~Gyr, corresponds to only 0.07\% of the mass of Ceres. If the loss rate of H$_2$O in Ceres remained constant throughout its thermal history, the initial water reservoir is thus likely to be integraly preserved. Moreover, since the other volatile species are expected to be trapped as hydrates, clathrate hydrates and pure condensates in this reservoir, we can conclude that they have also been preserved from outgassing throughout the thermal history of the asteroid.

Ceres being the largest and, due to its size, probably the wettest Main Belt asteroid, it is an ideal target for carrying out observations aiming at constraining its water regime. The experiment searching for water being vaporised within the polar regions of Ceres should be repeated with the state-of-the-art instrumentation available today on large telescopes. Such a detection would confirm unambiguously the presence of a large amount of water near the surface of Ceres. Direct observation of water ice, or of the effects of hydration, on the surface of Ceres can also be attempted for lower latitudes on the asteroid using a combination of high-angular resolution and spectroscopic instruments permitting the full
resolution of its surface to the ~30-40 km level. Due to its low spectral resolution, imaging of the surface of Ceres, even when it is spatially resolved using HST or adaptive optics, is not sensitive to the presence of ice, while the detection of such is within the reach of low resolution spectroscopic observations (e.g. the detection of absorption features in the 1.0-3.5 $\mu$m region). A spatially resolved spectroscopic mapping of the surface of Ceres in the near-infrared can be done with today's ground-based telescopes and would permit the mapping of the strength of the 3 $\mu$m band, and allow the search for regions on the surface where interstitial water ice, or hydration features could be present, for instance at the location of cracks within the surface of Ceres, or the locations of deep impact craters. Indeed, recent HST (Thomas et al. 2005) and adaptive optics (Carry et al. 2007) imaging observations of Ceres revealed the
presence of large impact craters across its surface which have likely disrupted the outer crust of the asteroid enough to directly expose the sub-surface mantle of wetter material. Finally, a spectroscopic study of the surface of Ceres, in order to search for the spectral signature of water and maybe those of other volatiles, should not be limited to one wavelength region (although the near-infrared range offers many diagnostic bands) but should, instead, encompass a wider range, from the near-UV to infrared wavelengths, in order to improve the identification of the chemicals species responsible for these spectral features. 

Finally, the NASA Discovery mission {\it Dawn}, which has been launched in September 2007 and whose arrival at Ceres is
scheduled for 2015, will certainly bring new constraints on the
presence of volatiles in the Main Belt. In particular, the {\it Dawn}
mapping spectrometer (MS) covers the spectral range from the near UV
(0.25 \micron) through the near IR (5 \micron) and has moderate to
high spectral resolution and imaging capabilities (Russell et
al. 2004). These characteristics make it an appropriate instrument for
determining the asteroid's global surface composition. Near-infrared
mapping of the surface of Ceres at small spatial scales will
be very sensitive to volatile concentrations and may reveal ice spots
on fresh impact-crater ridges. Moreover, the gravitation investigation
of Ceres will allow the determination of its gravity field up
to the 12th harmonic degree (Russell et al. 2004). Such a measurement
will enable the shape and gravity models to characterize crustal and
mantle density variations and, consequently, the amount of volatiles
trapped therein.

\section*{Acknowledgments}

This work was supported in part by the Swiss National Science Foundation. J.H. gratefully acknowledges the financial support provided by PPARC. We thank Jean-Marc Petit and Jeffrey Cuzzi for helpful remarks. Many thanks to the anonymous Referee whose useful comments invited us to strengthen our manuscript.

\label{lastpage}


\begin{thebibliography}{99}

\bibitem[\protect\citeauthoryear{A'Hearn \& Feldman}{1992}]{1992Icar...98...54A} A'Hearn M.~F., Feldman P.~D., 1992, Icar, 98, 54

\bibitem[\protect\citeauthoryear{Alibert, Mordasini, \& Benz}{2004}]{2004A&A...417L..25A} Alibert Y., Mordasini C., Benz W., 2004, A\&A, 417, L25

\bibitem[\protect\citeauthoryear{Alibert et al.}{2005}]{2005a} Alibert Y., Mousis O., Benz W., 2005a, ApJ, 622, L145

\bibitem[\protect\citeauthoryear{Alibert et al.}{2005}]{2005b} Alibert Y., Mousis O., Mordasini, C., Benz W., 2005b, ApJ, 626, L57

\bibitem[\protect\citeauthoryear{Alibert, Mordasini, \& Benz}{2005}]{2005A&A...417L..25A} Alibert Y., Mordasini C., Benz W., Winisdoerffer, C.  2005c, A\&A, 434, 343

\bibitem[\protect\citeauthoryear{Allamandola et al.}{1999}]{1999SSRv...90..219A} Allamandola L.~J., Bernstein M.~P., Sandford S.~A., Walker R.~L., 1999, SSRv, 90, 219

\bibitem[\protect\citeauthoryear{Andreas}{2007}]{2007Icar..186...24A} Andreas E.~L., 2007, Icar, 186, 24

\bibitem[\protect\citeauthoryear{Britt et al.}{2002}]{2002aste.conf..485B} Britt D.~T., Yeomans D., Housen K., Consolmagno G., 2002, aste.conf, 485

\bibitem[\protect\citeauthoryear{Bus, Vilas, \& Barucci}{2002}]{2002aste.conf..169B} Bus S.~J., Vilas F., Barucci M.~A., 2002, aste.conf, 169

\bibitem[\protect\citeauthoryear{Busarev}{1998}]{1998Icar..131...32B} Busarev V.~V., 1998, Icar, 131, 32

\bibitem[\protect\citeauthoryear{Carry}{2007}]{Carry07} Carry, B., Dumas, C., Fulchignoni, M., Merline, W., Berthier, J., Hestroffer, D., Fusco, T., Tamblyn, P., 2007, A\&A, submitted

\bibitem[\protect\citeauthoryear{Cassen}{1994}]{1994Icar..112..405C} Cassen P., 1994, Icar, 112, 405

\bibitem[\protect\citeauthoryear{Cellino et al.}{2003}]{2003Icar..162..278C} Cellino A., Diolaiti E., Ragazzoni R., Hestroffer D., Tanga P., Ghedina A., 2003, Icar, 162, 278

\bibitem[\protect\citeauthoryear{Chevrier et al.}{2007}]{2007GeoRL..29x..45M} Chevrier V., Sears D.W.G., Chittenden J.D., Roe L.A., Ulrich R., Bryson K., Billingsley L., Hanley J., 2007, GeoRL, 34, 02223 

\bibitem[\protect\citeauthoryear{Chick \& Cassen}{1997}]{1997ApJ...477..398C} Chick K.~M., Cassen P., 1997, ApJ, 477, 398

\bibitem[\protect\citeauthoryear{Ciesla \& Cuzzi}{2006}]{2006Icar..181..178C} Ciesla F.~J., Cuzzi J.~N., 2006, Icar, 181, 178

\bibitem[\protect\citeauthoryear{Clark et al.}{2003}]{2003DPS....35.2205C} Clark B.~E., Rivkin A.~S., Bus S.~J., Sanders J., 2003, DPS, 35, 955

\bibitem[\protect\citeauthoryear{Clayton \& Mayeda}{1996}]{1996GeCoA..60.1999C} Clayton R.~N., Mayeda T.~K., 1996, GeCoA, 60, 1999

\bibitem[\protect\citeauthoryear{Cuzzi et al.}{2005}]{2005ASPC..341..732C} Cuzzi J.~N., Ciesla F.~J., Petaev M.~I., Krot A.~N., Scott E.~R.~D., Weidenschilling S.~J., 2005, ASPC, 341, 732

\bibitem[\protect\citeauthoryear{Cyr, Sears, \& Lunine}{1998}]{1998Icar..135..537C} Cyr K.~E., Sears W.~D., Lunine J.~I., 1998, Icar, 135, 537

\bibitem[\protect\citeauthoryear{Dotto et al.}{2000}]{2000A&A...358.1133D} Dotto E., et al., 2000, A\&A, 358, 1133

\bibitem[\protect\citeauthoryear{Dullemond et al.}{2004}]{2004A&A...} Dullemond C.~P., Dominik, C., 2004, A\&A, 417, 159

\bibitem[\protect\citeauthoryear{Durda et al.}{2004}]{2004Icar..167..382D} Durda D.~D., Bottke W.~F., Enke B.~L., Merline W.~J., Asphaug E., Richardson D.~C., Leinhardt Z.~M., 2004, Icar, 167, 382

\bibitem[\protect\citeauthoryear{Durham, McKinnon, \& Stern}{2005}]{2005GeoRL..3218202D} Durham W.~B., McKinnon W.~B., Stern L.~A., 2005, GeoRL, 32, 18202

\bibitem[\protect\citeauthoryear{Fanale \&  Salvail}{1989}]{1989Icar...82...97F} Fanale F.~P., Salvail J.~R., 1989, Icar, 82, 97

\bibitem[\protect\citeauthoryear{Farmer}{1976}]{1976Icar...28..279F} Farmer C.~B., 1976, Icar, 28, 279

\bibitem[\protect\citeauthoryear{Fegley}{2000}]{2000SSRv...92..177F} Fegley B.~J., 2000, SSRv, 92, 177 

\bibitem[\protect\citeauthoryear{Fogg \& Nelson}{2006}]{2006IJAsB...5..199F} Fogg M.~J., Nelson R.~P., 2006, IJAsB, 5, 199

\bibitem[\protect\citeauthoryear{Gautier et al.}{2001}]{2001ApJ...559L.183G} Gautier D., Hersant F., Mousis O., Lunine J.~I., 2001a, ApJ, 559, L183
 
\bibitem[\protect\citeauthoryear{Gautier et al.}{2001}]{2001ApJ...550L.227G} Gautier D., Hersant F., Mousis O., Lunine J.~I., 2001b, ApJ, 550, L227

\bibitem[\protect\citeauthoryear{Gibb et al.}{2004}]{2004ApJS..151...35G} Gibb E.~L., Whittet D.~C.~B., Boogert A.~C.~A., Tielens A.~G.~G.~M., 2004, ApJS, 151, 35

\bibitem[\protect\citeauthoryear{Gomes et al.}{2005}]{2005Natur.435..466G} Gomes R., Levison H.~F., Tsiganis K., Morbidelli A., 2005, Natur, 435, 466

\bibitem[\protect\citeauthoryear{Hardersen, Gaffey, \& Abell}{2005}]{2005Icar..175..141H} Hardersen P.~S., Gaffey M.~J., Abell P.~A., 2005, Icar, 175, 141

\bibitem[\protect\citeauthoryear{Hersant, Gautier, \& Hur{\'e}}{2001}]{2001ApJ...554..391H} Hersant F., Gautier D., Hur{\'e} J.-M., 2001, ApJ, 554, 391 

\bibitem[\protect\citeauthoryear{Hersant, Gautier, \& Lunine}{2004}]{2004P&SS...52..623H} Hersant F., Gautier D., Lunine J.~I., 2004, P\&SS, 52, 623

\bibitem[\protect\citeauthoryear{Hilton}{2002}]{2002aste.conf..103H} Hilton J.~L., 2002, aste.conf, 103

\bibitem[\protect\citeauthoryear{Hsieh, Jewitt, \& Fern{\'a}ndez}{2004}]{2004AJ....127.2997H} Hsieh H.~H., Jewitt D.~C., Fern{\'a}ndez Y.~R., 2004, AJ, 127, 2997

\bibitem[\protect\citeauthoryear{Hsieh \& Jewitt}{2006}]{2006Sci...312..561H} Hsieh H.~H., Jewitt D., 2006, Sci, 312, 561

\bibitem[\protect\citeauthoryear{Irvine \& Knacke}{1989}]{1989oeps.book....3I} Irvine W.~M., Knacke R.~F., 1989, oeps.book, 3

\bibitem[\protect\citeauthoryear{Ivan}{2007}]{Ivan.book....3I} Ivantsov, A., 2007, P\&SS, submitted

\bibitem[\protect\citeauthoryear{Jewitt et al.}{2007}]{2007prpl.conf..863J} Jewitt D., Chizmadia L., Grimm R., Prialnik D., 2007, prpl.conf, 863

\bibitem[\protect\citeauthoryear{King et al.}{1992}]{1992Sci...255.1551K} King T.~V.~V., Clark R.~N., Calvin W.~M., Sherman D.~M., Brown R.~H., 1992, Sci, 255, 1551

\bibitem[\protect\citeauthoryear{Kochetova}{2004}]{2004SoSyR..38...66K} Kochetova O.~M., 2004, SoSyR, 38, 66

\bibitem[\protect\citeauthoryear{Kouchi et al.}{1994}]{1994A&A...290.1009K} Kouchi A., Yamamoto T., Kozasa T., Kuroda T., Greenberg J.~M., 1994, A\&A, 290, 1009

\bibitem[\protect\citeauthoryear{Krasinsky et al.}{2002}]{2002Icar..158...98K} Krasinsky G.~A., Pitjeva E.~V., Vasilyev M.~V., Yagudina E.~I., 2002, Icar, 158, 98 

\bibitem[\protect\citeauthoryear{Kress \& Tielens}{2001}]{2001M&PS...36...75K} Kress M.~E., Tielens A.~G.~G.~M., 2001, M\&PS, 36, 75 

\bibitem[\protect\citeauthoryear{Kuzmanoski \& Kova{\v c}evi{\'c}}{2002}]{2002A&A...395L..17K} Kuzmanoski M., Kova{\v c}evi{\'c} A., 2002, A\&A, 395, L17

\bibitem[\protect\citeauthoryear{Lazzarin et al.}{2004}]{2004A&A...425L..25L} Lazzarin M., Marchi S., Magrin S., Barbieri C., 2004, A\&A, 425, L25

\bibitem[\protect\citeauthoryear{Lewis \& Prinn}{1980}]{1980ApJ...238..357L} Lewis J.~S., Prinn R.~G., 1980, ApJ, 238, 357

\bibitem[\protect\citeauthoryear{Lebofsky et al.}{1981}]{1981Icar...48..453L} Lebofsky L.~A., Feierberg M.~A., Tokunaga A.~T., Larson H.~P., Johnson J.~R., 1981, Icar, 48, 453

\bibitem[\protect\citeauthoryear{Lide}{2002}]{2002crc..book.....L} Lide D.~R., 2002, crc..book,

\bibitem[\protect\citeauthoryear{Lodders}{2003}]{2003ApJ...591.1220L} Lodders K., 2003, ApJ, 591, 1220

\bibitem[\protect\citeauthoryear{Lunine \& Stevenson}{1985}]{1985ApJS...58..493L} Lunine J.~I., Stevenson D.~J., 1985, ApJS, 58, 493

\bibitem[\protect\citeauthoryear{Lupishko \& Belskaya}{1990}]{1990acm..proc..129L} Lupishko D.~F., Belskaya I.~N., 1990, acm..proc, 129

\bibitem[\protect\citeauthoryear{Lupishko}{2006}]{2005KFNTS...5..448L} Lupishko D.~F., 2006, SoSyR, 40, 214

\bibitem[\protect\citeauthoryear{McCord \& Sotin}{2005}]{2005JGRE..11005009M} McCord T.~B., Sotin C., 2005, JGRE, 110, 5009

\bibitem[\protect\citeauthoryear{Marchis et al.}{2003}]{2003Icar..165..112M} Marchis F., Descamps P., Hestroffer D., Berthier J., Vachier F., Boccaletti A., de Pater I., Gavel D., 2003, Icar, 165, 112

\bibitem[\protect\citeauthoryear{Marchis et al.}{2005}]{2005Icar..185...39M} Marchis, F., Descamps, P., Hestroffer, D., Berthier, J., 2005, Natur, 436, 822

\bibitem[\protect\citeauthoryear{Marchis et al.}{2006}]{2006Icar..185...39M} Marchis F., Kaasalainen M., Hom E.~F.~Y., Berthier J., Enriquez J., Hestroffer D., Le Mignant D., de Pater I., 2006, Icar, 185, 39

\bibitem[\protect\citeauthoryear{Matson et al.}{2007}]{2007Icar..187..569M} Matson D.~L., Castillo J.~C., Lunine J., Johnson T.~V., 2007, Icar, 187, 569

\bibitem[Mayer and Duschl (2005)]{Mayer06} Mayer, M. \& Duschl, W. J., 2005, MNRAS, 358, 614 

\bibitem[\protect\citeauthoryear{Michel et al.}{2001}]{2001Sci...294.1696M} Michel P., Benz W., Tanga P., Richardson D.~C., 2001, Sci, 294, 1696 

\bibitem[\protect\citeauthoryear{Michel, Benz, \& Richardson}{2004}]{2004P&SS...52.1109M} Michel P., Benz W., Richardson D.~C., 2004, P\&SS, 52, 1109

\bibitem[\protect\citeauthoryear{Mousis et al.}{2000}]{2000Icar..148..513M}  Mousis O., Gautier D., Bockel{\'e}e-Morvan D., Robert F., Dubrulle B., Drouart A., 2000, Icar, 148, 513

\bibitem[\protect\citeauthoryear{Mousis, Gautier, \& Bockel{\'e}e-Morvan}{2002}]{2002Icar..156..162M} Mousis O., Gautier D., Bockel{\'e}e-Morvan D., 2002, Icar, 156, 162

\bibitem[\protect\citeauthoryear{Mousis \& Gautier}{2004}]{2004P&SS...52..361M} Mousis O., Gautier D., 2004, P\&SS, 52, 361

\bibitem[\protect\citeauthoryear{Mousis \& Alibert}{2005}]{2005MNRAS.358..188M} Mousis O., Alibert Y., 2005, MNRAS, 358, 188

\bibitem[\protect\citeauthoryear{Mousis \& Marboeuf}{2006}]{2006DPS....38.1314M} Mousis O., Marboeuf U., 2006, DPS, 38, \#13.14

\bibitem[\protect\citeauthoryear{Mousis et al.}{2007}]{2007A&A...466L...9M}  Mousis O., Petit J.-M., Wurm G., Krauss O., Alibert Y., Horner J., 2007, A\&A, 466, L9

\bibitem[\protect\citeauthoryear{Nesvorn{\'y} et al.}{2006}]{2006Icar..183..296N} Nesvorn{\'y} D., Enke B.~L., Bottke W.~F., Durda D.~D., Asphaug E., Richardson D.~C., 2006, Icar, 183, 296

\bibitem[\protect\citeauthoryear{Novikov}{1969}]{Novikov} Novikov, P. A., Vagner, E. A., 1969, JEPTER, 17, 1377

\bibitem[\protect\citeauthoryear{Owen et al.}{1999}]{1999Natur.402..269O} Owen T., Mahaffy P., Niemann H.~B., Atreya S., Donahue T., Bar-Nun A., de Pater I., 1999, Natur, 402, 269

\bibitem[\protect\citeauthoryear{Pasek et al.}{2005}]{2005Icar..175....1P} Pasek M.~A., Milsom J.~A., Ciesla F.~J., Lauretta D.~S., Sharp C.~M., Lunine J.~I., 2005, Icar, 175, 1

\bibitem[\protect\citeauthoryear{Patashnick \& Rupprecht}{1975}]{1975ApJ...197L..79P} Patashnick H., Rupprecht G., 1975, ApJ, 197, L79

\bibitem[\protect\citeauthoryear{Porco \& Team}{2006}]{2006AGUFM.P22B..01P} Porco C., Team C., 2006, AGUFM, 1

\bibitem[\protect\citeauthoryear{Rivkin et al.}{2000}]{2000Icar..145..351R} Rivkin A.~S., Howell E.~S., Lebofsky L.~A., Clark B.~E., Britt D.~T., 2000, Icar, 145, 351

\bibitem[\protect\citeauthoryear{Russell et al.}{2004}]{2004P&SS...52..465R} Russell C.~T., et al., 2004, P\&SS, 52, 465

\bibitem[\protect\citeauthoryear{Sekine et al.}{2005}]{2005Icar..178..154S} Sekine Y., Sugita S., Shido T., Yamamoto T., Iwasawa Y., Kadono T., Matsui T., 2005, Icar, 178, 154

\bibitem[\protect\citeauthoryear{Shakura \& Sunyaev}{1973}]{1973A&A....24..337S} Shakura N.~I., Sunyaev R.~A., 1973, A\&A, 24, 337

\bibitem[\protect\citeauthoryear{Supulver \& Lin}{2000}]{2000Icar..146..525S} Supulver K.~D., Lin D.~N.~C., 2000, Icar, 146, 525

\bibitem[\protect\citeauthoryear{Thomas et al.}{2005}]{2005Natur.437..224T} Thomas P.~C., Parker J.~W., McFadden L.~A., Russell C.~T., Stern S.~A., Sykes M.~V., Young E.~F., 2005, Natur, 437, 224

\bibitem[\protect\citeauthoryear{Veras \& Armitage}{2004}]{2004MNRAS.347..613V} Veras D., Armitage P.~J., 2004, MNRAS, 347, 613 

\bibitem[\protect\citeauthoryear{Vernazza et al.}{2005}]{2005A&A...436.1113V} Vernazza P., Moth{\'e}-Diniz T., Barucci M.~A., Birlan M., Carvano J.~M., Strazzulla G., Fulchignoni M., Migliorini A., 2005, A\&A, 436, 1113

\bibitem[\protect\citeauthoryear{Veverka et al.}{1997}]{1997Sci...278.2109V} Veverka J., et al., 1997, Sci, 278, 2109

\bibitem[\protect\citeauthoryear{Weidenschilling}{1977}]{1977MNRAS.180...57W} Weidenschilling S.~J., 1977, MNRAS, 180, 57 

\bibitem[\protect\citeauthoryear{Wilson et al.}{1999}]{1999M&PS...34..541W} Wilson L., Keil K., Browning L.~B., Krot A.~N., Bourcier W., 1999, M\&PS, 34, 479

\end{thebibliography}
\end{document}